\documentclass[runningheads,a4paper]{llncs}

\usepackage[normalem]{ulem}
\usepackage{placeins}
\usepackage{color}
\usepackage[usenames,dvipsnames]{xcolor}
\usepackage{url,graphicx,times}
\usepackage{tabularx}
\usepackage{tabulary}
\usepackage{multirow}
\usepackage{booktabs}
\usepackage[square,comma,numbers,sort&compress,sectionbib]{natbib}
\usepackage{enumerate}
\usepackage{amsmath}
\usepackage{amssymb}
\usepackage{hyperref}
\usepackage{graphicx}
\usepackage{hyperref}
\usepackage[ruled,vlined,linesnumbered]{algorithm2e}
\usepackage{cleveref}
\usepackage{textcomp}
\usepackage{array}

\newcolumntype{K}[1]{>{\centering\arraybackslash}p{#1}}

\begin{document}

\mainmatter
\title{Towards an Understanding of Entity-Oriented Search Intents}

\author{Dar\'{i}o Garigliotti \and Krisztian Balog}
\institute{University of Stavanger, Stavanger, Norway
\email{\{dario.garigliotti,krisztian.balog\}@uis.no}}

\maketitle

\begin{abstract}
Entity-oriented search deals with a wide variety of information needs, from displaying direct answers to interacting with services.  In this work, we aim to understand what are prominent entity-oriented search intents and how they can be fulfilled.  We develop a scheme of entity intent categories, and use them to annotate a sample of queries.  Specifically, we annotate unique query refiners on the level of entity types.  We observe that, on average, over half of those refiners seek to interact with a service, while over a quarter of the refiners search for information that may be looked up in a knowledge base.  
\end{abstract}

\section{Introduction}
\label{sec:intro}

A large portion of information needs in web search look for specific entities~\citep{Pound:2010:AOR}.  
Entities are natural units for organizing information, and can provide not only more focused responses, but often immediate answers~\citep{Mika:2013:ESW}.  
Another type of entity-bearing queries is more transaction-oriented.
Either trying to book a flight or looking for tickets for an concert, just to mention two popular examples, users are often engaged to fulfill information needs by interacting with a third-party service or application.  
There has been an increasing focus on supporting task-based search~\citep{Kelly:2013:NWT}, and on modeling actionable knowledge; see, e.g., the dedicated vocabulary for actions in the schema.org ontology, and the NTCIR AKG task.\footnote{http://ntcirakg.github.io/tasks.html}  These developments display the interest and efforts
towards transforming search engines into actions-guided task completion assistants~\citep{Balog:2015:TCE}.
In this work, we are interested in studying one particular type of information needs, namely, entity-oriented searches.  Specifically, we want to answer a question arising from this web landscape: \emph{what do entity-oriented queries ask for?}  
Furthermore, which of those searches can be fulfilled by looking up direct answers from a knowledge base, and which would require to interact with external services?  

Most entity-oriented queries consist of an entity name, complemented with context terms, i.e., \emph{refiners}, to express the underlying intent of the user~\citep{Pound:2010:AOR}.  Examples of these queries are \emph{``the rock movies''} and \emph{``london book a hotel.''}  
Our main objective is to understand entity-related search intents by studying those refiners.  
Specifically, we represent refiners on the level of entity types.  Just like entity types boost the disambiguation of known entities and the grouping of emerging ones~\citep{Nakashole:2013:FST}, these type-level characterizations of entity refiners would favor knowledge abstraction and generalization.  
As an example, by representing with \emph{[city]} any entity of the type city, we want to categorize a refiner, e.g., ``rentals'', in the type-level query \emph{``[city] rentals''}.  
Then, we categorize these type-level refiners using an intent classification scheme. Our classification scheme comprises four main categories: property, website, service, and other.  

We perform this study without having direct access to past usage data or query logs.  
To overcome the absence of such data, we exploit query suggestions from a major search engine API. This strategy has been employed successfully in previous work for various applications~\citep{Fourney:2011:CUI,Benetka:2017:AIN}.
After acquiring query suggestions for entities of a given type, they are aggregated to extract type-level refiners.  Then, for a representative sample of 50 Freebase types, we collect human annotations for those refiners with respect to the classification scheme we developed.  

Our main findings show that, on average, more than a half of all unique type-level refiners correspond to interacting with external services, while over a quarter of them look for information that may be looked up in a knowledge base.  
Another contribution of this work is a large collection of type-level refiners, annotated with intent categories. The resources developed within this paper are made available at  \url{http://bit.ly/ecir2018-intents}.

\section{Related Work}

\citeauthor{Broder:2002:TWS}'s categorization of information needs is broadly accepted and is the most commonly used one for web search~\citep{Broder:2002:TWS}, with further refinements, e.g., in~\citep{Rose:2004:UUG,Jansen:2008:DIN}.  
We strive for a similar high-level categorization of intents, but specifically for entity-oriented search queries.  
Previous work has identified high-level patterns from web search queries.  For example, according to \citet{Lin:2012:AOA}, a query can be classified as an entity, an entity plus a refiner (e.g., \emph{``emma stone 2017''}), a category, a category plus a refiner (e.g., \emph{``doctors in barcelona''}), a website, or other sort of query.  
Such classification relies merely on lexico-syntactic forms and lacks a more semantically-grounded distinction.  

Search intents have been studied in previous work.   \citet{Reinanda:2015:MRR} explore entity aspects in user interaction log data.  Beyond finding aspects by comparing clustering methods over refiners, they address the tasks of ranking the intents for a given entity independently from a query and recommending aspects.  Unlike them, we (i) operate with individual query refiners (i.e., without clustering them together), (ii) model entity intents at the level of types, (iii) always consider entities in queries, and (iv) perform our study in the absence of search logs.

\section{Approach}
\label{sec:intents}

This section describes the process we followed for understanding entity-oriented search intents.  
An \emph{entity-oriented} or \emph{entity-bearing query} is a query that consists of an entity name possibly complemented with a refiner, usually as a suffix.  
Here, by \emph{entity} we mean an individual with its own independent existence, uniquely identified in a knowledge base~\citep{Balog:2017:ER}.  
More than just a syntactic complement, a \emph{refiner} is a complementary surface form expressing an underlying user \emph{intent} in relation with the entity.  
As an example, consider the entity \emph{keens steakhouse} (a restaurant) in the search query \emph{``keens steakhouse menu.''}  The refiner ``menu'' expresses the intent of reading the restaurant's menu.  
To understand what these entity-bearing queries ask for, we characterize the refiners on the level of \emph{entity types}, where an entity type is a semantic class that groups entities together with common characteristics.  For example, one of the types of \emph{Albert Einstein} in Freebase is \texttt{award\_winner}.

Our approach, to be detailed in the next subsections, can be summarized as follows.  
We collect refiners for a set of prominent entities, and aggregate them across entity types to obtain type-level refiners.  
Next, we develop a classification scheme of \emph{intent categories}, with a focus on how to fulfill the intent expressed by a type-level refiner.  
Finally, we annotate a representative sample of entity types with intent categories, and obtain a corpus of prominent type-level refiners assigned to those categories.

\subsection{Collecting Refiners}
\label{sec:intents:id}

We use the type system of Freebase.  It is a two-layer categorization system, where types on the leaf level are grouped under high-level domains.  
Specifically, we use the latest public Freebase dump (2015-03-31), discarding domains meant for administering the Freebase service itself (e.g., \texttt{base}, \texttt{common}).

We focus on prominent entities, since in this way we benefit from observing a larger and more representative selection of information needs.  As the criterion of an entity prominence, we rely on Wikistats page views.\footnote{\url{https://dumps.wikimedia.org/other/pagecounts-ez/}}  This dataset registers the number of times its English Wikipedia article has been requested.   
We set empirically a prominence threshold of 3,000~page views per article over a span of one year (from June 2015 to May 2016).  Given a Freebase type, we select it if it covers at least 100 entities with a prominence above the threshold.  Applying these criteria, the selected set contains 634~types.  

In a second step, we collect query suggestions from the Google Suggestions API for at most top 1,000~entities per type according to the above prominence criteria.  Then, we replace the name of the entity by its type in each query suggestion.  This can be viewed as getting queries where a refiner complements the type.  For example, the type-level query \emph{``[travel destination] map''} is obtained from all queries for popular travel destinations, e.g., \emph{``sydney map''} and \emph{``paris map.''}  
Finally, we retain only those refiners that occur in at least 5~suggestions for the given type.  This leads to a total of 2,688~distinct type-level refiners for 631~types.

\subsection{Classification Scheme}
\label{sec:intents:scheme}

To address our main goal of understanding entity-related search intents, we need a suitable scheme to classify the entity intents.
After a close inspection of the type-level refiners, we define the following scheme of \emph{intent categories}.  These categories are focused on how (and from which type of source) the information need can be fulfilled.

\begin{itemize}
	\item \textbf{Property}: The refiner looks for a specific entity property or attribute that can be looked up in a knowledge base.  For example, ``children'' in the query \emph{``angelina jolie children``} or ``opening times'' in \emph{``at\&t stadium opening times.''}
	\item \textbf{Website}: The refiner is about reaching a specific website or application.  For example, ``twitter'' in the query \emph{``karpathy twitter.''}  This category is a rough equivalent of navigational queries in~\citep{Broder:2002:TWS}. 
	\item \textbf{Service}: The refiner expresses the need to interact with a service, possibly by redirecting to an external site or app. For example, ``menu'' in the query \emph{``keens steakhouse menu''} would indicate the need for  accessing to an external site for reading the restaurant's menu.  As another example,  ``new album'' in \emph{``eric clapton new album''} looks for a service to read about, or listen to, or buy the new album.
	\item \textbf{Other}: None of the previous ones is applicable.  For example, ``batman'' in the query \emph{``christian bale batman''} serves to disambiguate the person's role of interest.
\end{itemize}

\subsection{Annotation}
\label{sec:intents:annot}

We need to sample a set of representative types, since it is unfeasible to annotate all types in the knowledge base.
From the set of 631 types, we perform stratified sampling  as follows.  We sort the types by the total aggregated frequencies of refiners.  We delimit 5 roughly equally-sized intervals by the splitting values of 1,500, 3,000, 6,000, and 8,500 refiners per type; we randomly pick 10 types from each interval.  We annotate data for this final set of 50 representative Freebase types.

We used crowdsourcing to annotate type-level refiners with intent categories.  
Specifically, using the Crowdflower platform, for each annotation instance we presented workers with the query, indicating its entity type and refiner, and asked them to select one of the four intent categories.  A total of 5,301 unique instances (type-level refiners) were annotated, each by at least 3 judges (5 at most, if necessary to reach a majority agreement, using dynamic judgments).  We paid \textcent 5 per batch, comprising 11 annotation instances.  
We ensured quality by requiring a minimum accuracy of 80\%, a minimum time of 20 seconds per batch, and a minimum confidence threshold of 0.7.
For each type, we only retain an annotated refiner if at least three annotators agreed on the majority category.  This leads to a total of 2,313 unique refiners.

\section{Results and Analysis}
\label{sec:results}

\begin{figure}[!t]
	\centering
	\includegraphics[width=0.32\textwidth]{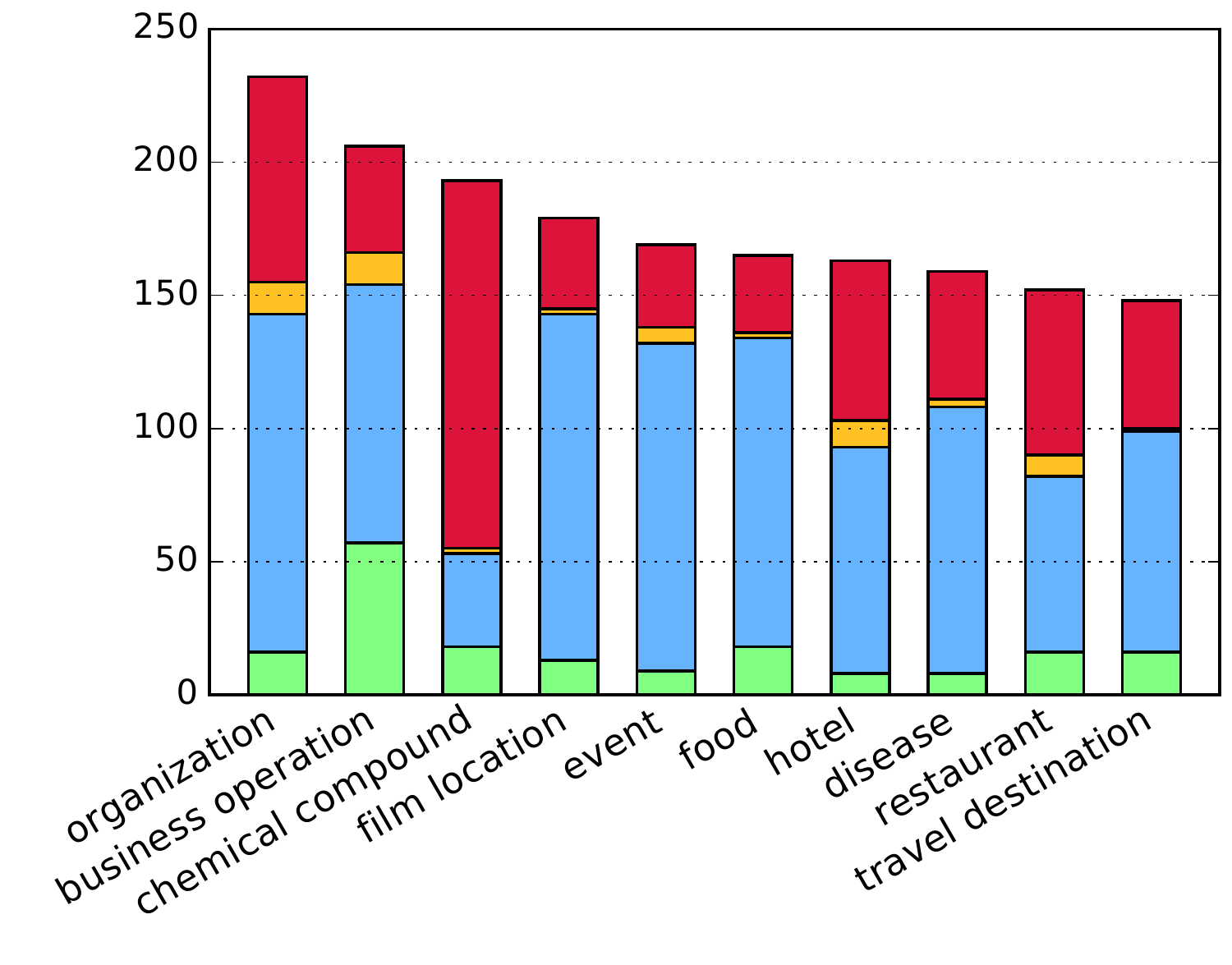}
	\includegraphics[width=0.32\textwidth]{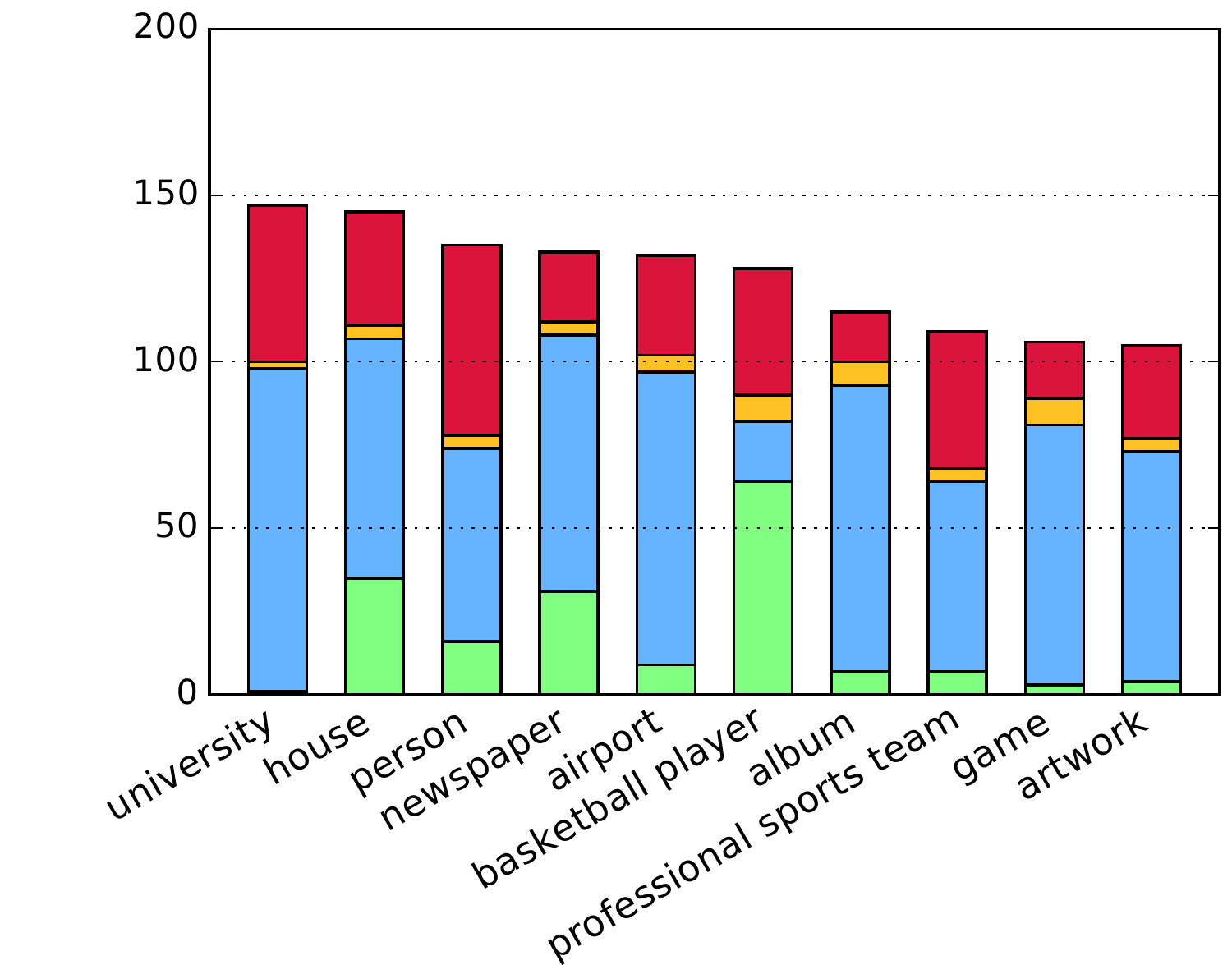}
	\includegraphics[width=0.32\textwidth]{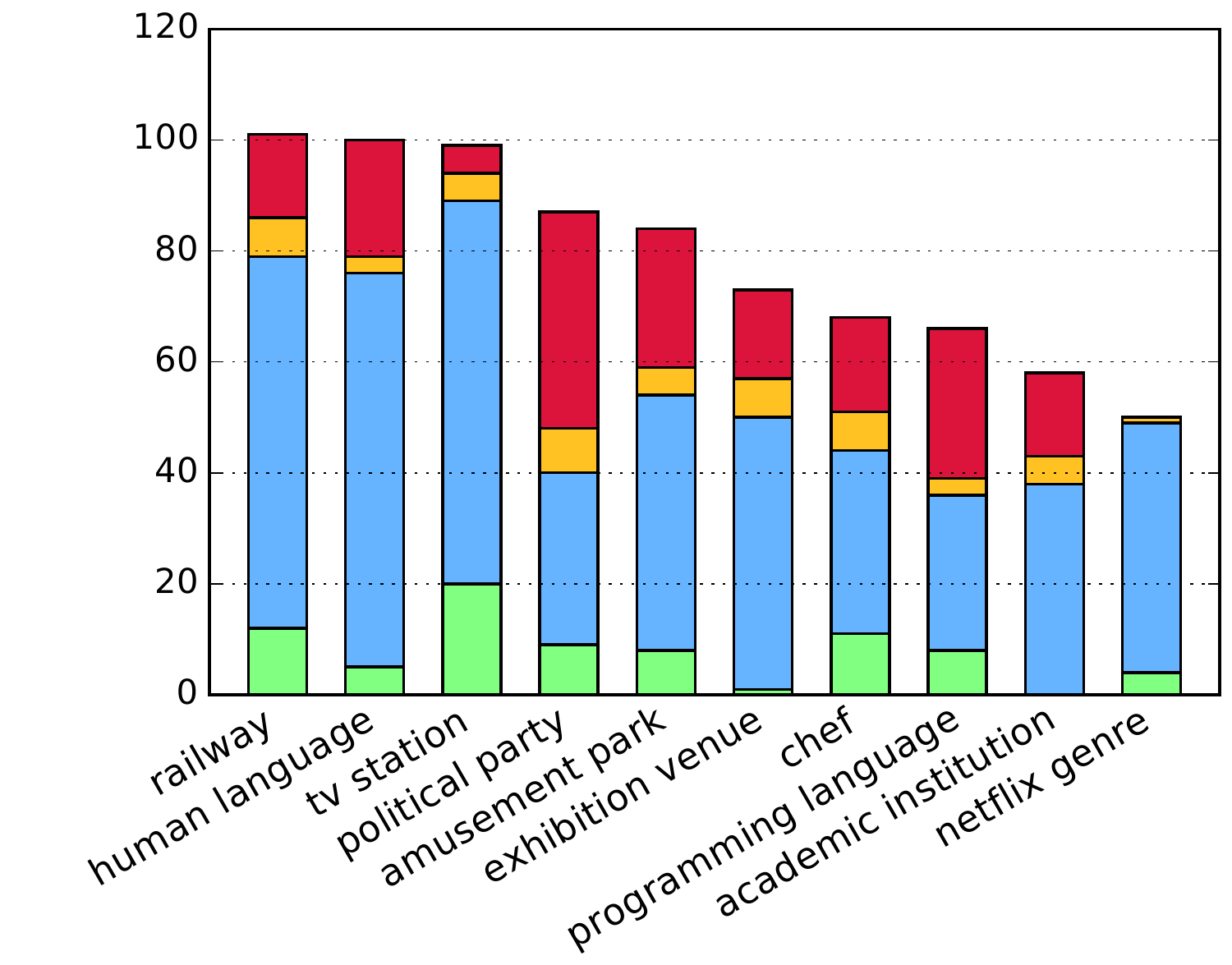}
	\includegraphics[width=0.32\textwidth]{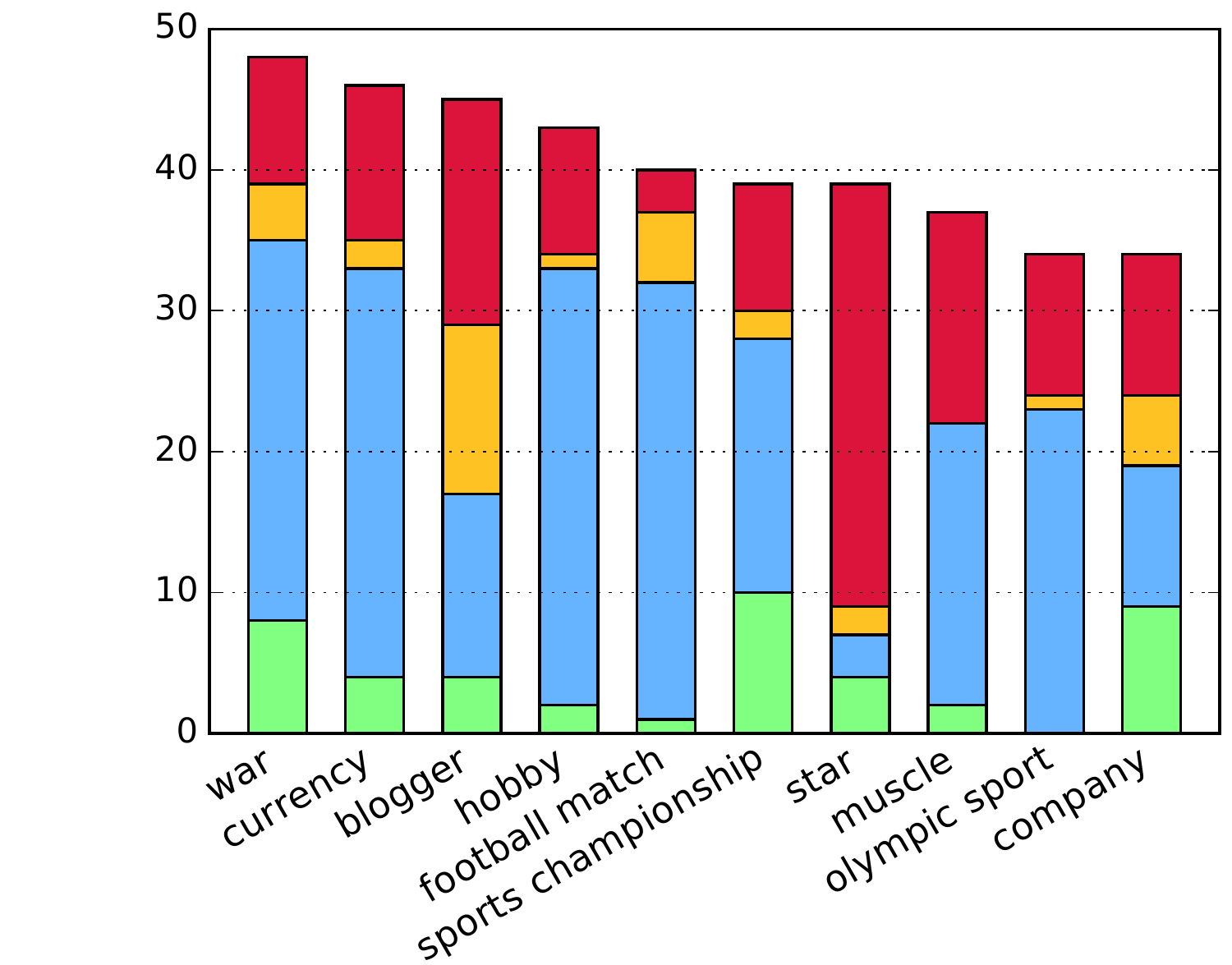}
	\includegraphics[width=0.32\textwidth]{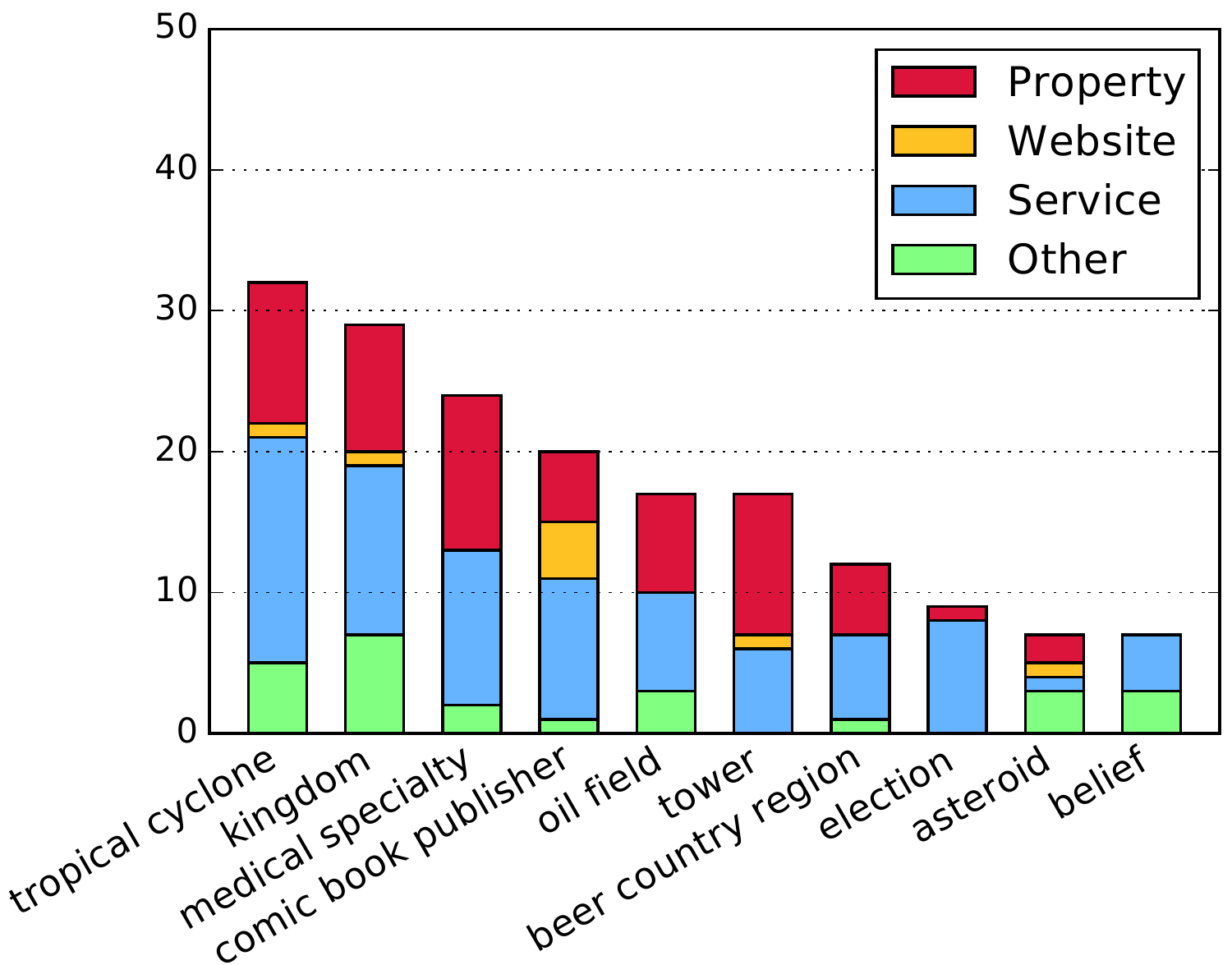}
	\caption{Distributions of intent categories for the sampled types.  Note that the y-axis scales differ.}
	\label{fig:exper-types_vs_qs_x5}
\end{figure}

 Figure~\ref{fig:exper-types_vs_qs_x5} presents the number of refiners classified per each category, for the 50 sampled types, grouped in one plot for each of the 5 intervals of the stratified sampling.  Since the final set of types was sampled from types with prominent entities, this ordering, given by the number of refiners, in a way also reflects the prominence of types.

We obtain a distribution of entity intent categories per type after normalizing the frequency of each category by the total of refiners for that type.  
From the average proportions in these distributions, we can answer our initial questions.  A 54.06\% of unique entity-oriented queries are to be fulfilled by interacting with some external service or app, meanwhile, 28.6\% look for direct answers from a knowledge base.  Further, 5.34\% of the type-level refiners represent an attempt to reach a website, while 12.08\% of them do not fit into any of the previous three categories.

The types with the largest proportion of \emph{service} intents are \texttt{netflix\_genre} (with refiners, e.g., ``videos,'' ``live''), \texttt{election} (``map,'' ``polls''), \texttt{football\_match} (``vi\-deo,'' ``highlights''), and \texttt{music\_album}.    
The \emph{property} intent category covers refiners that are of a more static nature, e.g.,  
\texttt{chemical compound} (with refiners like ``structural formula,'' ``molecular weight''), \texttt{political\_party} (``slogan,'' ``president''), \texttt{star} (``type of star,'' ``temperature''), or \texttt{tower} (``hours,'' ``height''); only the first one is a very prominent type.
Most of the entity types exhibit a non-empty proportion of \emph{website} intents.  
Among all the types, this category exceeds the average proportion, e.g., for \texttt{organization},
\texttt{business\_operation}, \texttt{hotel} and \texttt{blogger}.    
The most frequent website refiners in the whole corpus are ``wikipedia,'' ``twitter,'' ``facebook,'' and ``youtube.''  
For a few types like \texttt{muscle}, \texttt{election}, \texttt{belief}, or \texttt{medical\_speci\-ality}, all in the lowest populated groups, no website refiner is present.
A marginal proportion of refiners are classified as having the \emph{other} intent.    
A few exceptional cases with large proportions of other intents are, e.g., \texttt{business\_operation} and \texttt{house} (where the refiner is usually a location), or \texttt{basketball\_player} (for which many refiners refer mostly to an NBA franchise, e.g., ``lakers'').
Table~\ref{table:analysis:examples} provides additional examples for a selection of types.

\begin{table}[!h]
  \centering
  \caption{Examples of refiners for each intent category, for each (stratified) type group.} 
  \label{table:analysis:examples}
    \scriptsize
  \begin{tabular}{p{3cm} p{2cm} p{2.6cm} p{2.6cm} p{1.7cm} }
    \toprule
    Entity type & \multicolumn{4}{c}{Intent category} \\    
    \cline{2-5}
    & Property & Website & Service & Other \\
    \midrule
  	\texttt{comic\_book\_publisher} & logo, address & wiki, website, twitter & submissions, publishing, & movies \\
  	& & & comics & \\
  	\texttt{tower} & height, address, & wiki & tickets, restaurant & collapse \\
  	& opening hours &  &  & \\

    \midrule
 	\texttt{war} & deaths, results, & youtube, wikipedia, & video, uniforms, & ap euro, \\
 	& cause & reddit, quizlet & pictures, documentary & in hindi \\
  	\texttt{academic\_institution} & logo, email, & wiki, login, & scholarships, ranking, & baseball \\
  	& notable alumni & twitter, portal & map, library, jobs & \\
    \midrule
    \texttt{automotive\_company} & stock, logo, & wikipedia, website, & parts, careers, & india, inc \\
    & ceo, address & linkedin, facebook & investor relations & \\
    \texttt{programming\_language} & syntax, ide & wikipedia, & jobs, examples, & 3, 2017 \\
    & & wiki, github & interview questions & \\
    \midrule
    \texttt{restaurant} & phone number, & yelp, twitter, app, & wine list, vouchers, & sf, nj, nyc \\
    & owner, location & tripadvisor, groupon & recipes, menu prices & \\
    \texttt{music\_album} & value, cast, & youtube, wikipedia, & zip download, video, & 2015, lp \\
    & release date & amazon, imdb & ukulele chords, tracklist & \\
    \midrule
	\texttt{person} & son, salary, & youtube, instagram, & tour, quotes, & sr, now, \\
	& real name & snapchat & photos, new album & ww2 \\
	\texttt{travel\_destination} & zip code, & craigslist & weather radar, vacation, & today, nj \\
	 & train station &  & tours, things to do & \\
    \bottomrule
  \end{tabular}
  \vspace{-0.05in}  
\end{table}
\normalsize
%

\section{Conclusions and Future Work}
\label{sec:concl}

The study performed in this work has lead to a better understanding of what entity-oriented queries ask for.  We have developed a classification scheme to categorize entity-oriented search intents and annotated a representative sample of type-level refiners using this scheme.  
We have found that, on average, more than a half of those are to be fulfilled by interaction with services; another large proportion of information needs look for direct answers from a knowledge base.  
Several lines of future work arise from our study.  One of them is to develop a method for automatic intent categorization.    
Another direction is the clustering of refiners which express the same underlying intent.  
Finally, we seek to extend our approach to be able to capture tail entities and intents.

\FloatBarrier

\renewcommand*{\bibfont}{\scriptsize}
\bibliographystyle{abbrvnat}
\bibliography{ecir2018-actions}

\end{document}